\documentclass[12pt,preprint]{aastex}

\shorttitle{MID INFRARED OBSERVATIONS OF VAN MAANEN 2}
\shortauthors{J. Farihi}

\begin{document}

\title{MID INFRARED OBSERVATIONS OF VAN MAANEN 2:\\
	NO SUBSTELLAR COMPANION}

\author{J. Farihi and E. E. Becklin}
\affil{Department of Physics \& Astronomy, 8371 Math Sciences Building, 
	University of California, Los Angeles, CA 90095-1562}
\email{jfarihi@astro.ucla.edu,becklin@astro.ucla.edu}

\and

\author{B. A. Macintosh}
\affil{Institute of Geophysics \& Planetary Physics, Lawrence
 	Livermore National Laboratory, 7000 East Avenue L-413,
	Livermore, CA 94551}
\email{bmac@igpp.llnl.gov}

\begin{abstract}

The results of a comprehensive infrared imaging search for
the putative $0.06M_{\odot}$ astrometric companion to the
4.4 pc white dwarf van Mannen 2 are reported.  Adaptive optics
images acquired at $3.8\mu$m reveal a diffraction limited core
of $0.09''$ and no direct evidence of a secondary.  Models predict
that at 5 Gyr, a $50M_{\rm J}$ brown dwarf would be only 1 magnitude
fainter than van Maanen 2 at this wavelength and the astrometric
analysis suggested a separation of $0.2''$.  In the case of a chance
alignment along the line of sight, a 0.4 mag excess should be
measured.  An independent photometric observation at the same
wavelength reveals no excess.  In addition, there exist published
{\em ISO} observations of van Maanen 2 at $6.8\mu$m and $15.0
\mu$m which are consistent with photospheric flux of a 6750 K white
dwarf.  If recent brown dwarf models are correct, there is no
substellar companion with $T_{\rm eff}\ga500$ K.

\end{abstract}


\section{INTRODUCTION}

Van Maanen's star, also known as van Maanen 2 was discovered
in 1917 by Adrian van Maanen \citep{van17}.  Its discovery was
quite accidental, as van Maanen was searching for common proper
motion companions to HD4628 (Lalande 1299) and noticed the even
larger proper motion of a nearby star.  At 4.4 pc, van Maanen 2
is the nearest single white dwarf \citep{hol02}.

\citet{mak04} reported the astrometric detection of a substellar
companion to van Maanen 2 through analysis of {\em Hipparcos} data.
The data suggested an orbital solution with a lower bound companion
mass of $0.06\pm0.02M_{\odot}$, a period of 1.57 yr, and a maximum
separation on the sky of $0.3''$.  Based on the published parameters,
a binary orbital calculator (A. Ghez 2004, private communication)
indicated a separation of $0.19''$ at position angle $274\arcdeg$
near the middle of January 2004.  This paper presents the results
of a direct imaging search carried out during that time frame.

\section{OBSERVATIONS, DATA REDUCTION, \& PHOTOMETRY}

\subsection{\it Keck Adaptive Optics Observations}

Van Maanen 2 was observed using the facility adaptive optics system 
\citep{win00a,win00b} and the NIRC2 camera at Keck Observatory on
12 Jaunuary 2004.  Five dithered images were obtained through an
$L^{'}$ ($3.4-4.1\mu$m) filter.  To reduce thermal backgrounds
and provide a more symmetric PSF, a circular pupil stop was used
which describes an inscribed circle on the hexagonal Keck primary,
with a diameter of $\sim9$ meters.  Each image consisted of 100 coadds
of 0.4 seconds each.  Van Maanen 2 itself was used as the guide
star, with the adaptive optics system running at a rate of 80 Hz.
The correction was quite good, with a Strehl ratio (measured
relative to the PSF of the 9 meter pupil) of 0.75 and a full width
at half maximum of $0.089''$, essentially diffraction limited.  The
plate scale was $0.01''$ per pixel.

The adaptive optics images were reduced using standard
programs in the IRAF environment.  For a given raw image,
the sky was extracted by taking the median of the four remaining
dithered images with rejection of hot and cold pixels.  A flat
frame was created by averaging all five sky frames and subsequently
normalizing.  Each dithered frame was then sky subtracted, flat
fielded, and the resulting five frames were registered and averaged.

Due to the uncertainties in the wings of the adaptive optics
PSF and weather that was not photometric, it was difficult to
perform photometry on the Keck data.  Therefore photometric
observations of van Maanen 2 were acquired at the IRTF.

\subsection{\it IRTF and ISO Observations}

$3.8\mu$m data on van Maanen 2 and two standard stars were
acquired on 1 February 2004 by Alan Stockton at the 3 meter
NASA Infrared Telescope Facility with NSFCAM \citep{ray03}.
For the standards, a 9 point dither pattern was used with each
frame consisting of 0.15 second exposure times 100 coadds -- 
yielding a total integration time of 135 seconds.  The UKIRT
faint standard SA92-342 ($L^{'}=10.44$ mag) was observed
immediately before van Maanen 2 and the Elias standard star
HD22686 ($L^{'}=7.19$ mag) was observed immediately afterward
\citep{leg03}.  Two sets of images, acquired in the same manner
as the standards, were obtained for van Maanen 2 for a total of
18 frames and 270 seconds integration.  Conditions during the
observations were reported as photometric.

The data on the two standards were reduced using
standard programs in the IRAF environment.  For a given raw
image, the sky was extracted by taking the median of the eight
remaining dithered images with rejection of hot and cold pixels.
A flat frame was created by averaging all nine sky frames and 
subsequently normalizing.  Each dithered frame was then sky
subtracted, flat fielded, and the resulting nine frames were
registered and averaged.  Reducing the two sets of images on van
Maanen 2 was more difficult due to the faintness of the target in
each raw frame.  It was found that the target was more easily
seen in pairwise subtracted images rather than in sky subtracted
images.  All 18 pairwise subtracted frames were flat fielded, 
registered and averaged, creating one final image for photometric
measurements.

The flux of both standards was measured with aperture radii
of $1.2''$ and $2.4''$ and corrected for extinction.  The error
in determining the zero point was 0.03 mag.  Van Maanen 2 was
measured in the smaller aperture to minimize noise, corrected
to the larger aperture (with an error of 0.04 mag) and extinction
corrected.  This yielded a signal to noise of 48 and a measurement
error of 0.03 mag.  The photometric measurement including all errors
was $L^{'}=11.40\pm0.06$ mag for van Maanen 2.

ISOCAM observations of van Maanen 2 were carried out in
1997 in an effort to provide observational constraints
on the origin of metals in the photospheres of cool white
dwarfs \citep{cha99}.  Data were taken at $6.8\mu$m and
$15.0\mu$m on van Maanen 2 and several other white dwarfs
as well as an A0V calibrator star.  The data are listed in
Table \ref{tbl-1}.

\section{RESULTS \& DISCUSSION}

\subsection{\it Adaptive Optics - No Direct Evidence}

In order to estimate the brightness of the reported companion,
a model for a $50M_{\rm J}$ brown dwarf at 5 Gyr was chosen.  This
mass lies conservatively in the lower range of possible values and
required no interpolation within available models.  The model age
is likely to be greater than the 3.67 Gyr cooling age \citep*{ber01}
of van Maanen 2 plus the $\sim0.5$ Gyr main sequence lifetime \citep
{mae89} for a $\sim4M_{\odot}$ progenitor of the $0.83M_{\odot}$ white
dwarf \citep{wei87,wei90,wei00,bra95,ber01}.  A substellar companion
of this mass and age would be a late T dwarf ($T_{\rm eff}\sim800$ K)
and have $M_{L^{'}}=14.3$ mag which is $L^{'}=12.5$ mag at 4.4 pc
\citep{bur97}.  Photometric $L^{'}$ band data on known brown dwarfs do
exist and the measurements agree with the models used here to within
0.3 mag for spectral type T6 \citep{leg02}.  Van Maanen 2 is predicted
to have $L^{'}=11.43$ mag based on the model predicted $V-K=0.84$ color
of a 6750 K helium white dwarf \citep{ber95} and the $K-L^{'}=0.12$
color of a 6750 K blackbody.  $V-K=0.89$ is the measured color -- the
extrapolation was done from $V$ in case of any contamination by a
companion at $K$.

\begin{figure}

\plotone{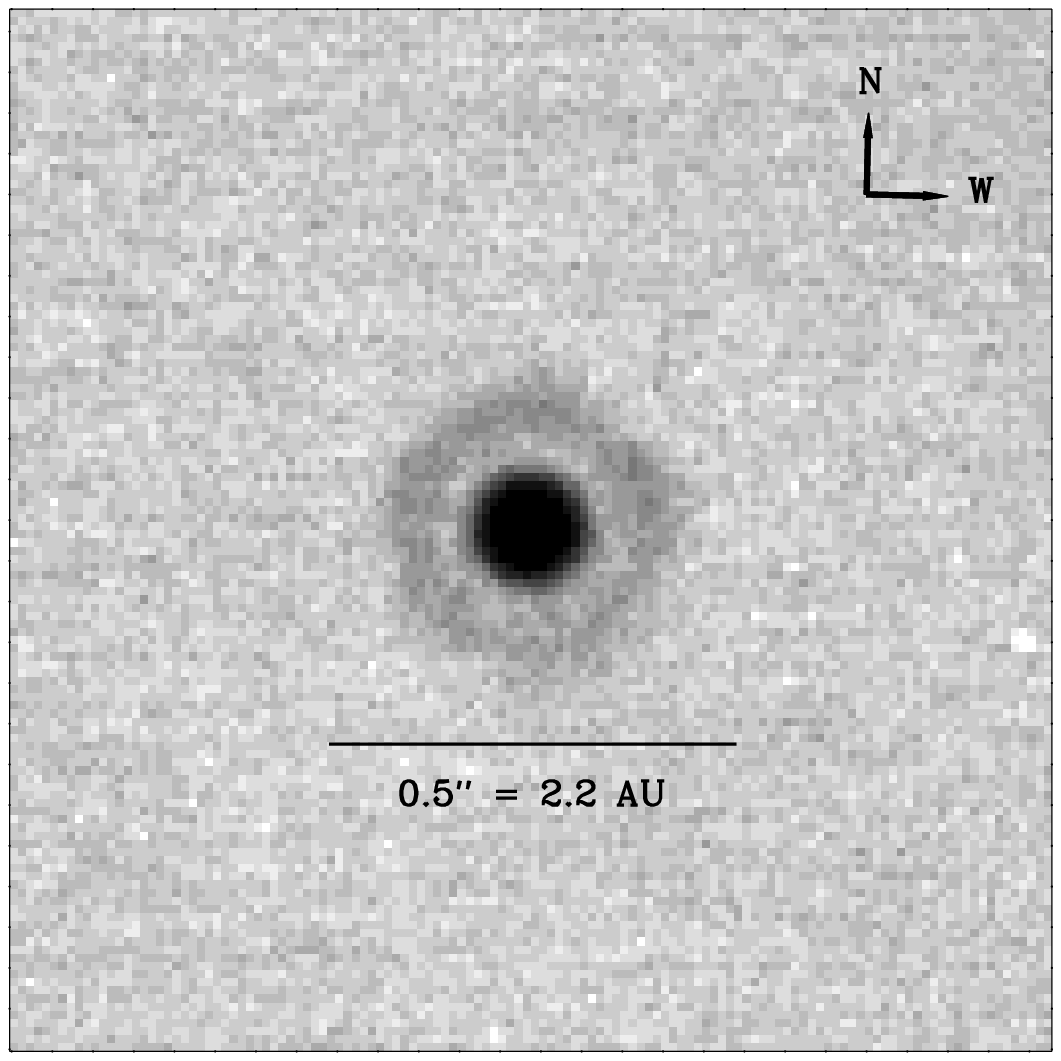}
\caption{$3.8\mu$m adaptive optics image of van Maanen 2.
There is no direct evidence for a companion one magnitude
fainter than the primary. \label{fig1}}
\end{figure}

The reduced $L^{'}$ image (Figure \ref{fig1}) shows no indication
whatsoever of a companion with the brightness expected from a
brown dwarf of the type reported by \citet{mak04}.  From the
published orbital parameters, the companion should have been at
a separation of $0.19''$ and position angle of $274\arcdeg$
on the date of the observation.  The full width at half maximum
of van Maanen 2 in the reduced image is $0.089''$ and the distance
to the first Airy ring is $\approx0.14''$.  There are two extremely
faint features within the Airy ring at position angles of 284 and
$295\arcdeg$.  Small aperture flux measurements, relative to the
primary, at eight different evenly spaced locations around the Airy
ring indicate that these features are unlikely to be real.  The flux
at 284 and $295\arcdeg$, $f/{f_0}=0.069$ and 0.073 respectively,
are both within $2\sigma$ (0.018) of the average flux ($f/{f_0}=
0.056$) in the ring and are almost certainly artifacts due to
imperfect optical guide star corrections.  If the feature at
$295\arcdeg$ were real, its brightness after subtracting the flux
of the primary in the Airy ring implies $L^{'}=15.9$ mag and a
mass of $\sim15M_{\rm J}$, assuming an age of 5 Gyr \citep{bur97}.
This is simply not massive enough to cause the reported
astrometric wobble.

\begin{figure}

\plotone{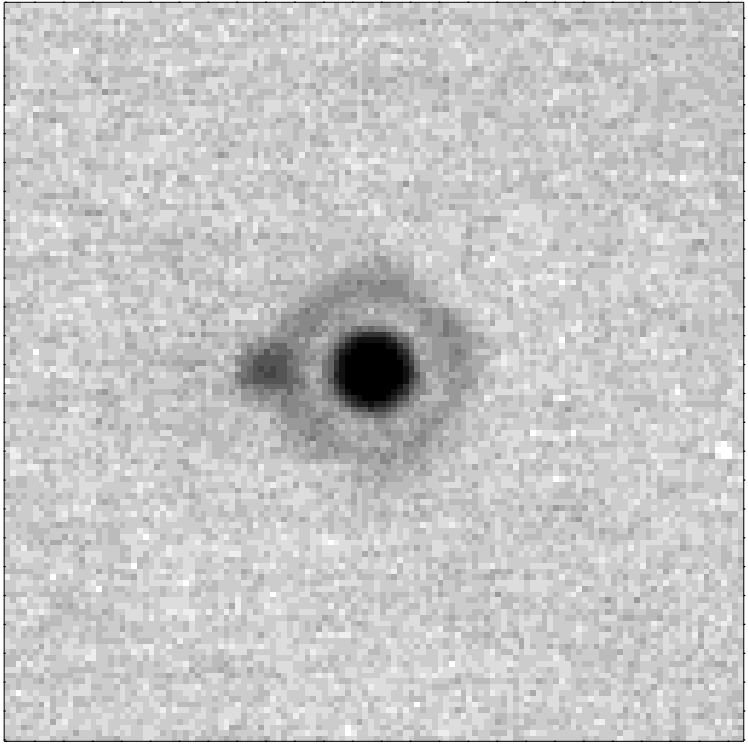}
\caption{Adaptive optics image of van Maanen 2 with an artificial
star placed at $0.19''$ from the white dwarf.  The magnitude of
the planted star is $L^{'}\approx13.4$ mag and is clearly seen.
\label{fig2}}
\end{figure}

An artificial star was planted at $0.19''$ from the primary
in order to test the ability to detect faint companions at this
separation.  The adaptive optics PSF, extracted from the reduced
image of van Maanen 2, was used for the stellar profile of the
planted star.  To be conservative, this artificial star is a full
2.0 magnitudes fainter than the white dwarf -- this was confirmed
by placing the star at many positions on the image and measuring
its flux without contamination by the primary.  The simulated star
is readily seen in Figure \ref{fig2}.  

Large errors in the reported orbital parameters allow for the
possibility that the companion remained unresolved due to chance
alignment during the observation.  The values and computed errors
in separation and position angle, derived from the binary orbital
calculator, are ${0.19''} \ ^{+0.07}_{-0.18}$ and ${274\arcdeg} \
^{+180}_{-2}$ respectively.  Hence the putative companion could
have been located at almost any position in its orbit by the
epoch of the observation.  However, there is further evidence
against this possibility.

\subsection{\it Photometry - No Indirect Evidence}

In the case of a chance alignment, any substellar companion
would still cause excess emission at mid infrared wavelengths.
This is primarily due to the size difference between brown dwarfs
and white dwarfs; a ratio of 10:1 in radius.  It is not possible
for a significant occultation to occur.  At most, a white dwarf
could block only $\sim1$\% of the light from an orbiting brown
dwarf.

As mentioned above, a value of $L^{'}=11.40\pm0.06$ mag was
measured for van Maanen 2.  This is to be compared with the
value predicted for a 6750 K white dwarf, $L^{'}=11.43$ mag.
The combined flux of the white dwarf plus a $50M_{\rm J}$ brown
dwarf at 5 Gyr would have a magnitude of $L^{'}=11.09$ mag.
Hence it is concluded there is no excess emission at this
wavelength.  

Furthermore, the published {\em ISO} observations place even
stronger constraints on the absence of flux from a brown dwarf
around van Maanen 2.  A recent model \citep*{bur03} of the flux
from a 5 Gyr, $25M_{\rm J}$ brown dwarf was integrated over the
{\em ISO} $15.0\mu$m filter LW3.  This results in a flux of about
1.0mJy in this filter.  The reported measurement taken at van Maanen
2 in 1997 was $0.5\pm0.2$ mJy.  This measurement is consistent
with photospheric flux from the white dwarf and an {\em excess} of
$5\sigma$ most likely would have been detected.  If the {\em ISO}
results are correct and the models are right, any companion with
$T_{\rm eff} \ga500$ K is ruled out -- this includes 10 Gyr old
brown dwarfs with $M\geq35M_{\rm J}$ \citep{bur97,bur03}.

\begin{figure}

\plotone{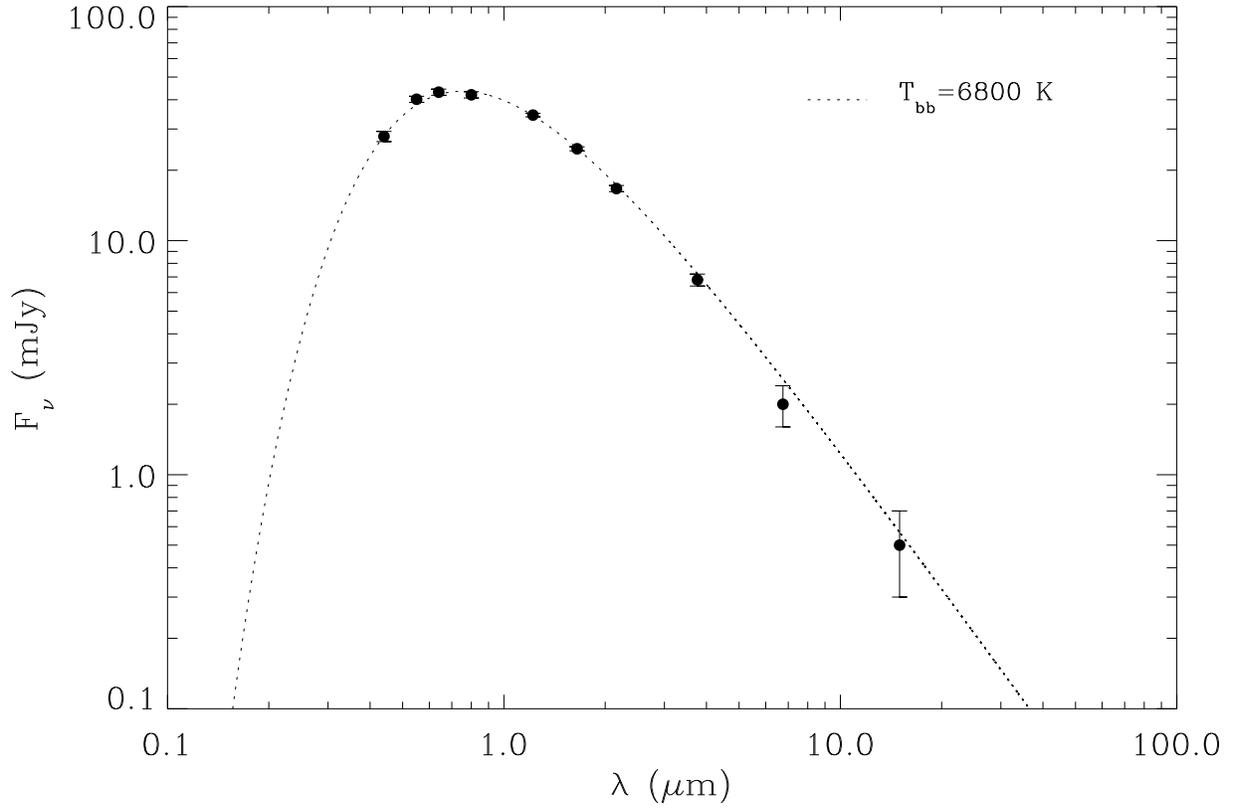}
\caption{Spectral energy distribution of van Maanen 2 from the
data in Table \ref{tbl-1}.  Overplotted is a 6800 K blackbody
demonstrating that the data reveal no excess emission in the infrared
and are consistent with a single star.
\label{fig3}}
\end{figure}

Table \ref{tbl-1} summarizes all existing photometric data
on van Maanen 2 and the corresponding fluxes are plotted in Figure
\ref{fig3}.  It should be clear from the figure that the measured
fluxes are all consistent with a single white dwarf with a
temperature around 6800 K.

\section{CONCLUSION}

Three compelling infrared observations of van Maanen 2 are
presented which rule out the presence of a substellar companion
warmer than $\sim500$ K.  First, $3.8\mu$m adaptive optics images
reveal no significant flux at the predicted magnitude and position
of the putative companion reported by \citet{mak04}.  Second,
$3.8\mu$m photometric observations are consistent with no excess
emission at this wavelength as would be expected from a $50M_
{\rm J}$ brown dwarf at 5 Gyr.  And third, $15.0\mu$m data also
are consistent with no excess emission, ruling out any possible
companion massive enough to induce the reported astrometric
wobble.

\acknowledgments

Part of the data presented herein were obtained at Keck
Observatory, which is operated as a scientific partnership
among the California Institute of Technology (CIT), the
University of California and the National Aeronautics and
Space Administration (NASA).  This publication makes use
of data acquired at the Infrared Telescope Facility, which
is operated by the University of Hawaii under Cooperative
Agreement no. NCC 5-538 with NASA, Office of Space Science,
Planetary Astronomy Program.  Some data used in this paper
are part of the Two Micron All Sky Survey (2MASS), a joint
project of the University of Massachusetts and the Infrared
Processing and Analysis Center (IPAC)/CIT, funded by NASA and
the National Science Foundation (NSF).  2MASS data were retrieved
from the NASA/IPAC Infrared Science Archive, which is operated
by the Jet Propulsion Laboratory, CIT, under contract with NASA.
The authors wish to express their gratitude to A. Stockton,
J. Rayner and A. Tokunaga for their assistance in acquiring
the IRTF observations of van Maanen 2.  Sincere thanks go to
J. Larkin and M. Barczys for donating some of their Keck adaptive
optics time for this project.  Acknowledgement also goes to B.
Hansen for bringing this system to our attention, to B. Zuckerman
for a careful reading of this manuscript and many discussions,
to A. Ghez for the binary orbit calculator, and to B. Vacca for
help in the data reduction.  Part of this work was performed under
the auspices of the U.S. Department of Energy, National Nuclear
Security Administration by the University of California, Lawrence
Livermore National Laboratory under contract No. W-7405-Eng-48.  
This research has been supported in part by grants from NASA to UCLA.

Facilities: \facility{Keck}, \facility{IRTF}

\clearpage

\clearpage

\clearpage

\clearpage

\begin{deluxetable}{lccr}
\tablecaption{Photometric Data on van Maanen 2 \label{tbl-1}}
\tablewidth{0pt}
\tablehead{
\colhead{Filter}			&
\colhead{$\lambda_{0}(\mu$m)} 		&
\colhead{Magnitude}			&
\colhead{$F_{\nu}$(mJy)}}

\startdata

$B$		& 0.44		& 12.91		& $27.9\pm1.4$\\
$V$		& 0.55		& 12.39		& $40.2\pm1.2$\\
$R$ 		& 0.64		& 12.13		& $43.1\pm1.3$\\
$I$ 		& 0.80		& 11.90		& $42.0\pm1.3$\\
$J$		& 1.22		& 11.69		& $34.4\pm0.6$ \\
$H$		& 1.65		& 11.57		& $24.7\pm0.5$ \\
$K_s$		& 2.16		& 11.50		& $16.7\pm0.5$ \\
$L^{'}$		& 3.76		& 11.40		& $6.8\pm0.4$  \\
LW2		& 6.75		& \nodata	& $2.0\pm0.4$  \\
LW3		& 15.0		& \nodata	& $0.5\pm0.2$  \\

\enddata

\tablecomments{$BVRI$ photometry is from \citet{ber01} and
is on the Johnson-Cousins system.  $JHK_s$ data are from 2MASS
and agree with the near infrared data reported by \citet{ber01}.
The $L^{'}$ data are from the present work and the LW2 \& LW3
fluxes are from {\em ISO} \citep{cha99}.}

\end{deluxetable}


\begin{thebibliography}{}

\bibitem[Bergeron et al.(2001)Bergeron, Leggett, \& Ruiz]{ber01}
 	Bergeron, P., Leggett, S. K., \& Ruiz, M. T. 2001, \apjs,
 	133, 413

\bibitem[Bergeron, Wesemael, \& Beauchamp(1995)]{ber95} Bergeron, P.,
        Wesemael, F., \& Beauchamp, A. 1995, \pasp, 107, 1047

\bibitem[Bessell \& Brett(1988)]{bes88} Bessell, M. S., \& Brett, J. M.
	1988, \pasp, 100, 1134

\bibitem[Bragaglia, Renzini \& Bergeron(1995)]{bra95} Bragaglia, A.,
        Renzini, A., \& Bergeron, P. 1995, \apj, 443, 735

\bibitem[Burrows et al.(1997)]{bur97} Burrows, A. 1997, \apj, 491, 856

\bibitem[Burrows et al.(2003)Burrows, Sudarsky, \& Lunine]{bur03} Burrows,
	A., Sudarsky, D., \& Lunine, J. I. 2003, \apj, 596, 587

\bibitem[Chary, Zuckerman, \& Becklin(1999)]{cha99} Chary, R., Zuckerman,
	B., \& Becklin, E.E. 1999, Conf. Proceedings: the Universe as Seen
 	by ISO, ed. P. Cox \& M. F. Kessler (Noordwijk: ESA/ESTEC), 289

\bibitem[Holberg, Oswalt, \& Sion(2002)]{hol02} Holberg, J. B., Oswalt,
	T. D., \& Sion, E. M. 2002, \apj, 571, 512

\bibitem[Leggett et al.(2002)]{leg02} Leggett, S. K. et al. 2002
	\apj, 564, 452

\bibitem[Leggett et al.(2003)]{leg03} Leggett, S. K. et al. 2003
	\mnras, 345, 144

\bibitem[Maeder(1989)]{mae89} Maeder, A. 1989, Proceedings of the
        ${\rm 5^{th}}$ IAP Workshop:  Astrophysical Ages and Dating
        Methods, eds. E. Vangioni-Flam, M. Cass\'e, J. Audouze, J.
        Tran Thanh Van, 71

\bibitem[Makarov(2004)]{mak04} Makarov, V. V. 2004, \apj, 600, L71

\bibitem[Rayner et al.(1993)]{ray03} Rayner, J. T., et al. 1993, SPIE,
	1946, 490

\bibitem[van Maanen(1917)]{van17} van Maanen, A. 1917, \pasp, 29, 258

\bibitem[Weidemann(1987)]{wei87} Weidemann, V. 1987, \aap, 188, 74

\bibitem[Weidemann(1990)]{wei90} Weidemann, V. 1990, \araa, 28, 103

\bibitem[Weidemann(2000)]{wei00} Weidemann, V. 2000, \aap, 363, 647

\bibitem[Wizinowich et al.(2000a)]{win00a} Wizinowich, P., et al.
	2000, \pasp, 112, 315

\bibitem[Wizinowich et al.(2000b)]{win00b} Wizinowich, P., Acton, D. S.,
 	Lai, O., Gathright, J., Lupton, W., \& Stomski, P., 2000, SPIE
 	4007, 64

\end{thebibliography}
\end{document}